\documentclass[aps,preprint,superscriptaddress,showpacs,pre]{revtex4}
\usepackage{graphicx}

\newcommand{\bd}[1]{\mbox{\boldmath$#1$}}

\begin{document}

\title{On nonlocally coupled complex Ginzburg-Landau equation}

\author{Dan Tanaka and Yoshiki Kuramoto}
\affiliation{Department of Physics, Graduate School of Sciences, 
Kyoto University, Kyoto 606-8502, Japan}

\begin{abstract}   
A Ginzburg-Landau type equation with nonlocal coupling 
is derived systematically 
as a reduced form of a universal class of reaction-diffusion systems
near the Hopf bifurcation point and in the presence of 
another small parameter.
The reaction-diffusion systems to be reduced are such that the chemical 
components constituting local oscillators are non-diffusive or 
hardly diffusive, 
so that the oscillators are almost uncoupled,
while there is an extra diffusive component which introduces effective 
nonlocal coupling over the oscillators.  
Linear stability analysis of the
reduced equation about the uniform 
oscillation is also carried out. This revealed that new types of instability 
which can never arise in the ordinary complex Ginzburg-Landau equation 
are possible, and their physical implication 
is briefly discussed. 
\end{abstract}

\maketitle

\section{introduction}
Oscillatory reaction-diffusion systems are generally
reduced to the complex Ginzburg-Landau (CGL) equation 
by means of the so-called center-manifold reduction when the local 
oscillators are close to their supercritical
Hopf bifurcation point\cite{Kura84}. This fact led to a now widely 
accepted view
that, without resorting to individual systems,
one may concentrate on CGL if one wishes to gain a qualitative
understanding of some universal dyanmical features
shared commonly by a broad class of 
oscillatory reaction-diffusion 
systems. This view in fact underlies a vast amount of work in the 
past (see \cite{CGLreview1,CGLreview2} for review) devoted to CGL.
However, due to its local nature of coupling, 
CGL may fail to capture some important aspects of the dynamics 
characteristic to 
a certain
class of oscillatory reaction-diffusion systems.   
In fact, it was argued recently that
a situation may arise where
the oscillator coupling becomes effectively nonlocal,
and as a consequence the system exhibits
such peculiar dynamics as can never be seen in CGL
\cite{Kura95,STC}.
This suggests that there remains
a yet
unexplored area of reaction-diffusion systems 
where nonlocal effects on the pattern dynamics must seriously
be considered. In the present paper, we are concerned with how this area
is still accessible within the framework of the center-manifold reduction.
It will turn out that this can actually be achieved
by a slight extension of the conventional reduction scheme.

Effective nonlocality in coupling may become relevant when the 
reaction-diffusion system involves
three or
more chemical components. Suppose that the system of concern
is such that the chemical
components constituting the local oscillators are free of diffusion, i.e.,  
the oscillators remain uncoupled, while the system involves an extra 
diffusive component 
which, for its diffusive nature,
plays the role of a coupling agent.
By eliminating mathematically this 
diffusive component,
the system becomes a field of nonlocally coupled oscillators possibly
involving delay also.
The main goal of the present paper is to achieve a 
reduction of such reaction-diffusion 
systems to a universal equation of the Ginzburg-Landau type
without missing the effects of nonlocality.

It may seem that the center-manifold idea would not work
for our purpose because CGL is believed to be quite a general result
of the center-manifold reduction. 
Still there seems to be a way out within the same framework
if we slightly extend
the conventional method of reduction.
Note first that under usual conditions the
disappearance of the effects of nonlocality near the bifurcation point
comes from that the characteristic wavelength $l_p$
of the pattern becomes far
longer than the effective coupling radius. If so, it may happen 
that nonlocality persists even close to the bifurcation point
when the system involves a certain parameter whose suitable 
tuning as the bifurcation point is approached
keeps $l_p$ comparable with the effective radius of
coupling.
Similar physical idea lies behind the multiple bifurcation theory
which aims to capture such complex dynamics as is absent in the vicinity 
of a 
simple bifurcation
point.

Section II starts with introducing a universal class of oscillatory
reaction-diffusion
systems involving three or more chemical components. 
After briefly discussing its physical relevance, we proceed to
its reduction near the Hopf bifurcation.
If a certain parameter associated with the strength of effective nonlocal
coupling is as small as the bifurcation parameter, the reduced equation
turns out to take the form of a nonlocally coupled complex
Ginzburg-Landau equation.
Analytic formulae for some coefficients of this equation are given, which
would be of great help in inferring possible ranges of the parameters
in the original system where nonlocality-dominated pattern dynamics is 
expected. Regarding the derivation of the reduced equation, 
our primary concern is the case when direct oscillator coupling is absent,
but the effects of weak diffusive (i.e.\ direct) coupling 
will also be considered. 
In Section III, linear stability analysis of 
our nonlocal CGL is carried out about the uniform oscillation. The 
resulting eigenvalue spectra, especially those of the phase branch,
can be qualitatively different from those of the standard CGL, whose
physical implication is discussed. A short summary will be given in
the final section.

\section{a universal class of reaction-diffusion systems and their 
reduction} 
The reaction-diffusion model of our concern 
was previously proposed by one of the present authors\cite{Kura95} and is
given by the general form
\begin{eqnarray}
\partial_t \bd{X} & = & 
\bd{f}(\bd{X}) + k \bd{g}(S), \\
\tau \partial_t S & = & -S + D \nabla^2 S + h(\bd{X}). 
\end{eqnarray}
Here the $n$-dimensional real vector field $\bd{f}$ represents a local 
limit-cycle oscillator with dynamical variable $\bd{X}$,
so that the first equation with $k=0$ represents 
a field of continuously distributed oscillators without
mutual coupling.
The full system involves an additional chemical 
component with concentration $S$ whose dynamics is 
governed by the second equation. 
This component simply diffuses and decays at a constant rate, while
it is produced
locally as represented by the term $h(\bd{X})$, 
the production rate depending on the local value of $\bd{X}$.
The dynamics of the local oscillators is influenced in return
by the local concentration of $S$, and
this effect is represented by the term $k\bd{g}(S)$.      
For the sake of convenience, we inserted in Eq.~(2) parameter $\tau$
to indicate explicitly the time constant of $S$, anticipating a limiting case
in which $\tau$ is vanishingly small. 
Note that there is no direct coupling among the 
local oscillators, while their indirect coupling is provided by the
diffusive chemical represented by $S$. In later discussions, we will
generalize the above model by assuming $\bd{g}$ and $h$ to depend both on
$\bd{X}$ and $S$
, and also by including
a small diffusion term i.e.\ weak direct coupling
in Eq.~(1).
Throughout the present paper, the spatial extension of the system is
supposed sufficiently large.

Let the spatially uniform steady state
of our system be 
given by $(\bd{X},S)=(0,0)$, or equivalently,
we measure $\bd{X}$ and $S$ always from their equilibrium values.
Thus, we clearly have $h(0)=0$.
It is also convenient to arrange $\bd{f}$ and $\bd{g}$ 
so that the equalities $\bd{f}(0)=\bd{g}(0)=0$ may be satisfied.
If we like, $S$ may be eliminated from the system by solving Eq.\ (2),
which can be done explicitly because the equation is linear. If the spatial
dimension is $d$, the solution
of Eq.~(2) is given by
\begin{eqnarray}
S(\bd{r},t)&=&(2\pi)^{-d}\int d\bd{q}\exp({i\bd{q}\bd{r}})\int _{-\infty}^{t}
\frac{dt'}{\tau} \nonumber\\
&&\exp\Bigl(-(1+Dq^{2})\frac{t-t'}{\tau}  \Bigr)
h_{\bd{q}}(t'),
\end{eqnarray}
where $h_{\bd{q}}(t')$ is the spatial Fourer transform of 
$h(\bd{X}(\bd{r},t'))$.
Equation (1) with $S$ given by Eq.~(3)
constitutes a field of nonlocally coupled
oscillators. Note that the nonlocality appears in time as well as in space,
and the characteristic scales in time and space associated with the
nonlocality are given by $\tau$ and $D^{1/2}$, respectively.
This fact will be used in the discussion below.

Suppose that $\bd{f}$ involves a parameter $\mu$ such that if $\mu<0$
each local system given by Eqs.~(1) and (2) with $D=0$
has a stable fixed point $(\bd{X},S)=(0,0)$
while this becomes oscillatory unstable
for $\mu\ge 0$; namely, at $\mu=0$ a pair of complex conjugate eigenvalues 
of the Jacobian
associated with each local system about the fixed point 
cross the imaginary axis of the complex plane, while
the other eigenvalues all remain in the left half plane.
It is a well known fact that the center-manifold reduction 
can be applied to reaction-diffusion systems 
near the Hopf bifurcation point of the local oscillators\cite{Kura84}. 
This leads 
generally to the complex Ginzburg-Landau (CGL) equation
\begin{equation}
\partial_{t}A=\mu\sigma A-\beta |A|^{2}A+\alpha\nabla^{2}A,
\end{equation}
where the amplitude $A$ and the parameters $\sigma$, $\alpha$ and $\beta$ are
generally complex. The small parameter $\mu$ has been retained 
in Eq.~(4) so that
$A$, $t$ and $\bd{r}$ scale like 
$|\mu|^{1/2}$, $|\mu|^{-3/2}$ and $|\mu|^{-1/2}$, respectively. 

It is clear that the above reduced form of reaction-diffusion systems
remains valid also
for our particular system given by 
Eqs.~(1) and (2).  
Since CGL is a diffusively coupled (i.e. locally coupled) system,
it may seem that 
effective nonlocality in coupling characteristic to our system
disappears completely near the bifurcation point. 
The reason for the disappearance of nonlocality
is clear. This is because the 
characteristic wavelength $l_p$ of the 
field $\bd{X}$
becomes longer and longer as the bifurcation point is approached like
$l_p\sim |\mu|^{-1/2}$  so that
the effective coupling radius given by $D^{1/2}$
comes to fall well within this scale, which gives nothing but the 
definition of local coupling. In what follows, 
we will be concerned with a special situation in which
spatial (and possibly temporal) nonlocality 
can survive even close to the bifurcation point,
so that
the reduced equation involves nonlocal coupling 
rather than
diffusive coupling. The same result was used already in earlier 
works\cite{Kura95,STC} for the particular case of vanishingly small $\tau$
without showing how the reduction can actually be achieved. We will develop
below the reduction procedure explicitly including the case of finite $\tau$.

Consider our system given by the form of Eqs.~(1) and (3) for which 
Eq.~(4) gives the right reduced form
near $\mu=0$ provided
there is no small parameter other than $\mu$. It is clear that the diffusion
term in Eq.~(4) is the reduced form of the coupling
term $k\bd{g}(S)$ in Eq.~(1). This implies that $|\alpha|=O(|k|)$.
Disappearance of spatial nonlocality is consistent with the fact that
the characteristic wavelength $l_p$ estimated from the dimensional argument
for Eq.~(4),
which confirms $l_p=O(|\mu/k|^{-1/2})$, 
is far larger than the effective coupling
radius given by $D^{1/2}$ provided $k$ remains an ordinary magnitude. 
This consistency is apparently broken 
if $k$ becomes as small as $\mu$
by which $l_p$ becomes 
independent of $\mu$ and hence can be comparable with the coupling radius. 
Putting it differently, spatial
nonlocality 
should remain relevant near the bifurcation point
provided the strength
of coupling 
between the local oscillators and the diffusive component becomes so weak as 
to satisfy
\begin{equation}
k\sim O(|\mu|).
\end{equation}
Thus, what we do next is to find a reduced equation
valid near the
doubly singular point $(\mu,k)=(0,0)$.
Before proceeding to this issue, however, 
we make a remark on {\em temporal} nonlocality.
Temporal nonlocality, which is  
characterized by the time constant $\tau$, 
may be relevant to the dynamics even near the 
bifurcation point.
However, this effect would never appear as a non-Markoffian
form of the reduced equation for the amplitude $A$ because the
time scale associated with the variation of 
$A$ is much longer than $\tau$. Still, $\tau$ could be comparable
with the period of the basic oscillation. Then, as we see below,  
the effect of delay may appear in the reduced equation through the
change in the space-dependence of the coupling function.
Specifically, when $\tau$ is finite,
the decay of the coupling function with distance becomes oscillatory
rather than monotone.

Suppose first that $k=0$. Then the reduced form of the ordinary equation (1)
is given by Eq.~(4) with $\alpha=0$. Although its derivation is routine, 
we now recapitulate it for the purpose of explaining soon later 
how easily the reduced equation can be
generalized when a small coupling term is introduced.

Let the Taylor expansion of $\bd{f}(\bd{X})$ in terms of $\bd{X}$ be
written as
\begin{equation}
\bd{f}(\bd{X})=L\bd{X}+M\bd{XX}+N\bd{XXX}+\cdots.
\end{equation}
Regarding the $\mu$-dependence of the coefficients
appearing in the above expansion,
we need to consider it only for the Jacobian $L$ to the first order in $\mu$
like $L=L_{0}+\mu L_{1}$; higher order corrections to $L$ as 
well as $\mu$-dependence of $M$, $N$ etc.~are irrelevant to the 
reduced equation to the leading order. 
Let the pure imaginary eigenvalues at $\mu=0$ be $\pm i\omega_0$, and
the corresponding right eigenvectors and its complex conjugate 
are written as $\bd{U}$ and $\bar{\bd{U}}$, respectively. 
The corresponding left eigenvectors and its complex conjugate
are denoted
by $\bd{U}^{*}$ and $\bar{\bd{U}}^{*}$, respectively. These eigenvectors 
satisfy
$\bd{U}^{*}\bar{\bd{U}}=\bar{\bd{U}}^{*}\bd{U}=0$ and $\bd{U}^{*}\bd{U}=
\bar{\bd{U}}^{*}\bar{\bd{U}}=1$.
If $\mu$ is nonvanishing but small,
the eigenvalues change to $\pm i\omega_{0}+\mu\lambda_{\pm}$, where 
$\bar{\lambda}_{+}=\lambda_{-}$.

To the lowest order in $\mu$, the original field $\bd{X}$ and the
complex amplitude $A$ are mutually related via
\begin{equation}
\bd{X}(t)={\rm e}^{i\omega_{0}t}\bd{U}A(t)+{\rm c.c.}.
\end{equation}
Thus, in this approximation we have $\partial_{t}A=
\exp(-i\omega_{0}t)\bd{U}^{*}(\dot{\bd{X}}-L_{0}\bd{X})$. 
This means further that the right-hand 
side of Eq.~(4) with $\alpha=0$
is identical with the reduced form of $\exp
(-i\omega_{0}t)\bd{U}^{*}(\bd{f}(\bd{X})-L_{0}\bd{X})$, or
\begin{eqnarray}
&&\mu\sigma A-\beta |A|^{2}A\simeq{\rm e}^
{-i\omega_{0}t}\bd{U}^{*}\Bigl(\bd{f}(\bd{X})-L_{0}\bd{X}\Bigr) \nonumber \\
&=&{\rm e}^{-i\omega_{0}t}\bd{U}^{*}(\mu L_{1}
\bd{X}+M\bd{XX}+N\bd{XXX}+\cdots).
\end{eqnarray}
The standard analysis determines the coefficients $\sigma$ and $\beta$
in terms of some parameters of the equations before reduction.
It is clear that the linear coefficient $\sigma$ is given by
\begin{equation}
\sigma=\bd{U}^{*}L_{1}\bd{U}
=\lambda_{+}.
\end{equation}
For obtaining $\beta$, the lowest order
expression for $\bd{X}$ given by Eq.~(7) is insufficient,
and we have to 
use a more precise formula including the next order term:
\begin{eqnarray}
\bd{X}(t)&=&{\rm e}^{i\omega_{0}t}\bd{U}A(t)+{\rm c.c.}\nonumber\\
&+&{\rm e}^{2i\omega_{0}t}\bd{V}_{+}A^{2}
+{\rm e}^{-2i\omega_{0}t}\bd{V}_{-}\bar{A}^{2}
+\bd{V}_{0}|A|^{2},
\end{eqnarray}
where
\begin{eqnarray}
\bd{V}_{+}&=&\bar{\bd{V}}_{-}=-(L_{0}-2i\omega_{0})^{-1}\bd{M}\bd{U}\bd{U},\\
\bd{V}_{0}&=&-2L_{0}^{-1}\bd{M}\bd{U}\bar{\bd{U}}.
\end{eqnarray}
Using these quantities, $\beta$ is given by
\begin{equation}
\beta=-2\bd{U}^{*}M\bd{UV}_{0}-2\bd{U}^{*}M\bar{\bd{U}}\bd{V}_{+}
-3\bd{U}^{*}N\bd{UU}\bar{\bd{U}}.
\end{equation} 

Suppose that the vector field $\bd{f}$ is modified slightly
to $\bd{f}+\bd{p}$.
It is clear that the corresponding reduced equation must also be modified
slightly with an additive term $\exp
(-i\omega_{0}t)\bd{U}^{*}\bd{p}$. The specific form of $\bd{p}$ of our concern
is the small coupling term $k\bd{g}(S)$
in Eq.~(1) with $S$ given by Eq.~(3). The original variable $\bd{X}$
and the reduced one $A$ must now be regarded as depending on space 
as well as on time.
Thus, our problem is to find a reduced form of the quantity
\begin{equation}
k\exp
(-i\omega_{0}t)\bd{U}^{*}\bd{g}(S)\equiv k\tilde{\bd{p}}
\end{equation}
using Eq.~(3).
Since $k$ is already small, we only need to consider the most dominant 
contribution to $\tilde{\bd{p}}$.
Noting that $\bd{g}(0)=0$, we may use a linear approximation 
\begin{eqnarray}
\bd{g}(S)&\simeq& \bd{g}_{0}S \nonumber \\
&\simeq&
\frac{\bd{g}_{0}}{(2\pi)^{d}}\int d\bd{q}\exp({i\bd{q}\bd{r}})\int _{-\infty}^{t}
\frac{dt'}{\tau} \nonumber\\
&&\exp\Bigl(-(1+Dq^{2})\frac{t-t'}{\tau}  \Bigr)
\bd{h}_{0}\bd{X}_{\bd{q}}(t'),
\end{eqnarray}
where $\bd{g}_{0}=d\bd{g}/dS|_{S=0}$ and 
$\bd{h}_{0}=dh(\bd{X})/d\bd{X}|_{\bd{X}=0}$.
Thus, using Eq.~(7) with $\bd{X}$ and $A$ supposed to depend also on
$\bd{r}$, we have
\begin{eqnarray}
\tilde{\bd{p}}
&=&\frac{\eta}{(2\pi)^{d}}\int d\bd{q}\exp({i\bd{q}\bd{r}})\int _{-\infty}^{t}
\frac{dt'}{\tau} \nonumber\\
\exp&&\!\!\!\!\!\!\!\!\!\!\!\Big(-(1+Dq^{2})\frac{t-t'}{\tau}-i\omega_{0}(t-t')  \Bigr)
A_{\bd{q}}(t'),    
\end{eqnarray}
where
\begin{equation}
\eta=(\bd{U}^{*}\bd{g}_{0})(\bd{h}_{0}\bd{U}).
\end{equation}
Note that Eq.~(16) ignores the contribution from the complex conjugate of
$A(t)$ which would give rise to a rapidly oscillating component of 
$\tilde{\bd{p}}$. This is allowed because such component would be averaged out
in the equation describing the slow evolution of $A(t)$. 
Since the time-integral in Eq.~(16) extends practically over the
interval between $t-\tau$ and $t$, the slowly varying amplitude $A_{\bd{q}}(t')$ may safely be replaced with $A_{\bd{q}}(t)$.
In this way, we obtain
\begin{equation}
k\tilde{\bd{p}}=k\eta' \int d\bd{r}'G(\bd{r}-\bd{r}')A(\bd{r}',t),
\end{equation}
where
\begin{equation}
G(\bd{r})=\frac{1}{(2\pi)^{d}}\int d\bd{q}
{\rm e}^{i\bd{q}\bd{r}}\frac{1+i\omega_{0}\tau}
{Dq^{2}+1+i\omega_{0}\tau},
\end{equation}
and 
\begin{equation}
\eta'=\frac{\eta}{1+i\omega_{0}\tau}.
\end{equation}
Note that $G(\bd{r})$ satisfies the normalization condition
\begin{equation}
\int d\bd{r}G(\bd{r})=1.
\end{equation}
Thus the final form of the reduced equation becomes
\begin{eqnarray}
\partial_{t}A&=&\mu\sigma A-\beta |A|^{2}A \nonumber\\
&&+k\eta' \int d\bd{r}'G(\bd{r}-\bd{r}')A(\bd{r}',t)
\end{eqnarray}
which we call nonlocally coupled complex Ginzburg-Landau 
equation or simply nonlocal CGL.
It is clear that the situation of interest is such that $k=O(|\mu|)$
for which the coupling term in the reduced equation is balanced in 
magnitude with the other terms even if the characteristic wavelength
is independent of $\mu$. 
We assume that the bifurcation is supercritical, i.e., the real part of 
$\beta$ is positive.

A few generalizations of the original system (Eqs.~(1) and (2)) can be made.
Firstly, $\bd{g}$ and $h$ may depend both on $\bd{X}$ and $S$.
Since the most dominant
part of these quantities alone is relevant to the reduced equation,
one may safely approximate $\bd{g}(\bd{X},S)$ and $h(\bd{X},S)$ as
$\bd{g}(\bd{X},S)=\bd{g}(\bd{X},0)+\bd{g}(0,S)$ and $h(\bd{X},S)=h(\bd{X},0)
+h(0,S)$, respectively. The resulting new term $k\bd{g}(\bd{X},0)$
may slightly modify $\bd{f}(\bd{X})$, but the modified $\bd{f}(\bd{X})$
may again be denoted by $\bd{f}(\bd{X})$. Furthermore, because $S$ is small, 
$h(0,S)$
is practically linear in $S$. Thus, this quantity simply modifies
the linear decay rate of $S$ which can be normalized by a suitable rescaling of
time. In this way, 
the final result of reduction
is unchanged except that $\bd{g}(S)$ and $h(\bd{X})$ are
replaced with $\bd{g}(0,S)$ and $h(\bd{X},0)$, respectively. As the second
generalization, we may include in Eq.~(1) a diffusion term like
\begin{equation}
\partial_t \bd{X} = 
\bd{f}(\bd{X}) +\hat{\delta}\nabla^{2}\bd{X}+k \bd{g}(S),
\end{equation}
where $\hat{\delta}$ 
is a diagonal diffusion matrix with non-negative elements. 
In parallel with the above argument for obtaining a reduced form of
the nonlocal coupling term, the reduced form of the quantity
$\exp
(-i\omega_{0}t)\bd{U}^{*}\hat{\delta}\nabla^{2}\bd{X}$
will then be added to the right-hand side of Eq.~(22).
To the lowest order approximation, one may apply Eq.~(7) for $\bd{X}$, by which
the above quantity becomes $\delta\nabla^{2}A$ where $\delta$ 
is a complex number with positive real part, and is
given by 
\begin{equation}
\delta=\bd{U}^{*}\hat{\delta}\bd{U}.
\end{equation}  
Thus, Eq.~(22) is modified as
\begin{eqnarray}
\partial_{t}A&=&\mu\sigma A-\beta |A|^{2}A+\delta\nabla^{2}A \nonumber\\
&&+k\eta' \int d\bd{r}'G(\bd{r}-\bd{r}')A(\bd{r}',t).
\end{eqnarray}
In the conventional 
reduction of reaction-diffusion systems, $|\delta|$ is assumed to be
of ordinary magnitude, so that the diffusion term can be
balanced with the linear and cubic terms in magnitude only if the 
characteristic
wavelength of $A$ scales like $|\mu|^{-1/2}$. However, the last property of
$A$ contradicts
with the particular situation of our concern
in which the characteristic
wavelength remains independent of $\mu$. 
Therefore, in what follows, we assume that $|\delta|$ as well as $k$ is of the
order of $\mu$, by which all terms on the right-hand side of Eq.~(25)
are balanced with each other, and the coupling nonlocality represented by the
last term can survive. 

It is more convenient to write Eq.~(25) in the form
\begin{eqnarray}
\partial_{t}A&=&\mu\sigma' A-\beta |A|^{2}A+\delta\nabla^{2}A \nonumber\\
&+&k\eta' \int d\bd{r}'G(\bd{r}-\bd{r}')\bigl(A(\bd{r}',t)-A(\bd{r},t)\bigr),
\end{eqnarray}
where $\sigma'=\sigma+\mu^{-1}k\eta'$. With this form, the coupling term
can be approximated by a diffusion term when the characteristic
wavelength of $A(\bd{r},t)$ is sufficiently longer than the coupling radius.
Note that $\sigma'$ remains ordinary magnitude because $k=O(\mu)$ by
assumption. Hereafter, we assume that the system is supercritical or
\begin{equation}
{\rm Re}\sigma'>0.
\end{equation}
   
An additional remark should be made on the functional form of the coupling
function $G$ in connection with the time scale $\tau$ of the diffusive 
component $S$. As implied by Eq.~(16), finite $\tau$ generally gives rise to
memory effects or temporal nonlocality after the variable $S$ has been
eliminated. However, as was noted before,
the reduced equation is free from memory effects because
the time scale of the slowly varying 
amplitude $A$ is much longer than $\tau$. We also noted that
because $\tau$ may be comparable
with the period $2\pi/\omega_{0}$ of the fundamental oscillation,
the effect of delay in coupling should be relevant to the reduced 
dynamics. Actually, as is clear from Eq.~(19), the effect of finite
$\tau$ appears in the coupling function $G$. For one-dimensional systems,
in particular, the coupling 
function is simply expressed as
\begin{equation}
G(x)=\frac{1}{2}(\alpha_{+}+i\alpha_{-}){\rm e}^
{-(\alpha_{+}+i\alpha_{-})|x|},
\end{equation} 
where 
\begin{equation}
\alpha_{\pm}=\sqrt{\frac{\pm 1+\sqrt{1+\theta^{2}}}{2D}}
\end{equation}
and $\theta=\omega_{0}\tau$.
If $\tau$ is vanishing, $G(x)$ decays exponentially with $|x|$, the decay 
length being given by $D^{1/2}$, whereas the decay becomes oscillatory when
$\tau$ is finite.
Thus the effect of delay in coupling is such that 
the complex amplitude $A(x')$ in the coupling term is multiplied by a factor 
$\exp(-i\alpha_{-}|x-x'|)$, or equivalently, the phase of $A(x',t)$, denoted 
by $\phi(x',t)$, is replaced with $\phi(x',t)-\alpha_{-}|x-x'|$.
In the physical language, this means that the phase 
at the spatial point $x'$ experienced by the oscillator at $x$ through the
delayed coupling cannot be its
current value $\phi(x',t)$ but should be the value at some time $t-t_0$ 
in the past, 
because
the phase information travels at a finite speed. 
If this speed is constant,
and the oscillation at $x'$ is nearly regular, then we have
$\phi(x',t-t_0)$ equal to $\phi(x',t)$ plus something proportional to
the distance $|x-x'|$, justifying the above result.

Coming back to general space dimensions, Eq.~(26) with the coupling function
given by Eq.~(19) involves many parameters. However, some of the parameters
can be eliminated by suitable
transformations of some variables. Firstly, the imaginary part of the
linear coefficient $\mu\sigma'$ vanishes through the transformation 
$A\rightarrow A\exp[(i\mu{\rm Im}\sigma')t]$. Secondly, one may rescale
$A$, $t$ and $\bd{r}$ in such a way that 
${\rm Re}\beta$, $\mu{\rm Re}\sigma'$ and $D$ may all become unity. 
In this way, we may write 
Eq.~(26) as  

\begin{eqnarray}
\partial_{t}A&=&A-(1+ic_{2})|A|^{2}A+(\delta_{1}+i\delta_{2})\nabla^{2}A 
\nonumber\\
+K(1\!\!\!&+&\!\!\!ic_{1})\int\!\! d\bd{r}
'G(\bd{r}-\bd{r}')\bigl(A(\bd{r}',t)-A(\bd{r},t)\bigr),
\end{eqnarray}
where the coupling function $G$ is defined as an integral form 
given by Eq.~(19), or
\begin{eqnarray}
G(\bd{r})&=&\frac{1}{(2\pi)^{d}}\int d\bd{q}
{\rm e}^{i\bd{q}\bd{r}}G_{q}, \\
G_{q}&\equiv&
\frac{1+i\theta}
{q^{2}+1+i\theta}.
\end{eqnarray}

The reduced equation now involves six independent parameters
$c_{1}$, $c_{2}$, $K$, $\theta$, $\delta_{1}$ and $\delta_{2}$ all of which are
independent of smallness parameters $\mu$, $k$ and $\delta$.
Note that the coupling coefficient in Eq.~(30) is related to some original 
parameters through
\begin{equation}
K(1+ic_{1})=\frac{k\eta'}{\mu{\rm Re}\sigma'}=\frac{k\eta'}{\mu{\rm Re}\sigma
+k{\rm Re}\eta'},
\end{equation}
or 
\begin{equation}
K=1-\frac{{\rm Re}\sigma}{{\rm Re}\sigma'}
\end{equation}
so that, by combining the last equation 
with the inequality (27) and the original
assumption ${\rm Re}\sigma>0$, we have a restrictive condition
\begin{equation}
K<1.
\end{equation}

\section{Eigenvalue spectrum about the uniform oscillation}
It is clear that Eq.~(30) admits a family of plane wave solutions
$A_{k}(\bd{r},t)=R\exp[i(\bd{kr}-\Omega t)]$ the
stability of which is extremely important to the 
understanding of the pattern dynamics of our system. 
In the present paper we will 
focus on the stability of the uniform oscillation $A_{0}(t)$ for which
\begin{eqnarray}
R&=&1,\\
\omega_{0}&=&-c_{2}.
\end{eqnarray}

We now put
\begin{equation}
A(\bd{r},t)=\Bigl(1+\Delta(\bd{r},t)\Bigr)A_{0}(\bd{r},t),
\end{equation}
and linearize Eq.~(30) in $\Delta(\bd{r},t)$.
The linearized equation can be solved in terms of the Fourier components
of $\Delta(\bd{r},t)$, denoted by $\Delta_{\bd{q}}(t)$.
Assuming its time-dependence in the form  $\Delta_{\bd{q}}(t)
\propto \exp(\lambda t)$,
we find the eigenvalue equation
\begin{equation}
\lambda^{2}+u(q)\lambda+v(q)=0,
\end{equation}  
where
\begin{eqnarray}
u(q)&=&-2{\rm Re}\gamma(q), \\
v(q)&=&|\gamma(q)|^{2}-|\gamma(0)|^{2},
\end{eqnarray}
with
\begin{eqnarray}
\gamma(q)&=&-(1+ic_{2})-(\delta_{1}+i\delta_{2})q^{2} \nonumber \\
&&+K(1+ic_{1})(G_{q}-G_{0}).
\end{eqnarray}
In what follows, we shall first 
concentrate on the case without diffusive coupling, i.e. 
$\delta_{1}=
\delta_{2}=0$; the effects of nonvanishing but 
small $\delta_1$ will be touched upon later.
 
Let the solutions of Eq.~(39) be denoted by
$\lambda_{+}$ and $\lambda_{-}$ with
${\rm Re}\lambda_{+}\ge {\rm Re}\lambda_{-}$. 
The uniform oscillation is stable if and only if ${\rm Re}\lambda_{+}<0$
holds for all $q$.
This holds only when $u(q)>0$ and $v(q)>0$ for all q. Equivalently, if
one of these 
inequalities becomes violated for a certain $q$, then the uniform oscillation
loses stability. This implies that the types of 
instability
could be classified in terms of the signs of $u(q)$ and $v(q)$.
Simple calculation shows that $u(q)$ and $v(q)$ can be expressed in the
following form where we use the notation $Q\equiv q^{2}$.
\begin{eqnarray}
u(q)&=&\xi(Q)\xi_{0}(Q), \\
v(q)&=&\zeta(Q)\zeta_{0}(Q)Q. 
\end{eqnarray} 
Here $\xi_{0}(Q)$ and $\zeta_{0}(Q)$
are non-negative functions of $Q$, and
\begin{eqnarray}
\xi(Q) &\equiv& a_2 Q^2+a_1 Q+a_0 , \\
\zeta(Q) &\equiv& b_1 Q+b_0 , \\
a_0 &=& 1+\theta^2 , \\
a_1 &=& 2+K(1+c_1\theta) , \\
a_2 &=& 1+K , \\
b_0 &=& 2K(1+c_1c_2+(c_1-c_2)\theta), \\
b_1 &=& K(2(1+c_1c_2)+K(1+c_1^2)).
\end{eqnarray}
Figure 1 shows the $a_{1}$-$a_{2}$ plane divided into 
three domains (labeled by A, B and C)
corresponding to qualitatively different 
forms of $u(q)$. Note that the $a_{1}$-$a_{2}$ plane 
covers all possible types of $u(q)$
because $a_{0}$ is positive definite.
Similar picture for $v(q)$ in the $b_{0}$-$b_{1}$ plane is 
displayed in Fig.~2
where the whole space is divided into four domains 
(labeled by A$'$ to D$'$). 
Since the parameters $a_1$, $a_2$, $b_0$ and $b_1$ as functions of
four independent parameters $K$, $\theta$, $c_1$ and $c_{2}$
can be changed independently,
every combination between two members, one from the group
(A,B,C) and the other from (A$'$,B$'$,C$'$,D$'$), is possible.
Uniform oscillation is unstable for all these combinations except for
(A,A$'$). Loss of stability occurs
as we move across the line $a_{2}=0$ or $a_{1}^2-4a_{0}a_{2}=0$ in Fig.~1,
for which the instability is oscillatory, 
or otherwise $b_{0}=0$ or $b_{1}=0$ in Fig.~2, for which the instability is
non-oscillatory. 
The critical wavenumber for which ${\rm Re}\lambda<0$ becomes first violated
equals zero on the line $b_{0}=0$, finite on
$a_{1}^2-4a_{0}a_{2}=0$ and infinite
on $a_{2}=0$ and $b_{1}=0$. The last type of instability,
i.e. the instability which starts at infinite $q$,
was called {\em short-wavelength bifurcation}
by Heagy et al.\cite{heagy}.

If $|K|$ is not too large, $u$ is non-negative, so that the instability 
is only through $v$ becoming negative and hence it is non-oscillatory.
The corresponding critical lines $b_{0}=0$ and $b_{1}=0$ in Fig.~2 
can now be translated
into critical relations between $K$ and $Kc_{1}$ under fixed 
$c_2$ and $\theta$. In this way, we have 
four types of eigenvalue
spectrum shown schematically in Fig.~3 where 
lines $L_1$ and $L_2$
correspond to $b_{0}=0$ and $b_{1}=0$, respectively. 
The figure shows the spectra only for phase-like
fluctuations, because in each case the amplitude branch which is separated from
the phase branch remains negative for all $q$ so that the amplitude-like 
fluctuations are irrelevant to stability. Separation between the phase and
amplitude branches also implies that the eigenvalues associated with these
branches are all real. In contrast, if these branches merge, then 
the eigenvalues
of the two branches for given $q$ would form a complex conjugate pair.
Note that in each type of spectrum shown in the figure 
the eigenvalue saturates to a constant as the wavenumber
tends to infinity, which is characteristic to nonlonally coupled
systems. The characteristic wavenumber about which the eigenvalue starts to
saturate equals the inverse of the coupling radius. Physically, this reflects
the fact that the dynamics of fluctuations whose wavelengths are much shorter
than the coupling radius is practically the same as the oscillators'
individual dynamics, and hence it is insensitive to the wavenumber. 
Note also that the type U$_2$ spectrum is possible only when $\theta$ is
nonvanishing or, equivalently, when 
the nonlocal coupling before reduction involves delay.  

For larger $|K|$, the phase and amplitude branches can merge, so that
oscillatory instability becomes possible. In particular, the
short-wavelength type instability i.e.\ the instability 
initiated by fluctuations
with {\em infinite} wavenumber can  occur across the line $a_2=0$ or $K=-1$. 
This
type actually appears in the stability diagram in Fig.~4 which is a 
global extension
of the diagram in Fig.~3. The values of the other parameters are the same as in Fig.~3, i.e.,
$c_2=1.0$ 
and $\theta=6.0$. Oscillatory instability initiated by 
fluctuations with {\em finite}
wavenumber,
which occurs when crossing the line $a_{1}^2-4a_{0}a_{2}=0$ in Fig.~1, does not
appear in Fig.~4, but may appear when different values of $c_2$ and $\theta$
are chosen.
For instance, the stability diagram for 
the case of $c_{2}=-2.5$ and $\theta=1.5$ is displayed in Fig.~5    
where the line $L_{4}$ gives the boundary associated with the last type of
instability.

Finally, we comment on the effects of weak diffusion represented by the
term $(\delta_{1}+i\delta_{2})\nabla^{2}A$ in Eq.~(30). 
For simplicity, we assume that $\delta_2$ is vanishing while
$\delta_1$ is a small positive.
It is clear that this diffusion term simply gives rise to an additional 
stabilizing 
term $-\delta_{1}q^2$ to each of the eigenvalues $\lambda_{\pm}$ as 
a function of $q$. Thus, the fluctuations with sufficiently short 
wavelength are always decaying. A particularly interesting result from
this fact appears in type U$_2$ eigenvalue spectrum in Fig.~3.
Depending on the value of $\delta_{1}$, the dispersion curve is deformed
like some curves shown in Fig.~6. Even the dispersion curves of types 
U$_1$ and U$_{12}$
could be deformed into a similar form if some parameter vaules are chosen
suitably.
It is clear that there is a critical
value of $\delta_{1}$ at which fluctuations with a certain wavenumber $q_c$
become unstable. Due to the presence of long-wavelength fluctuations which are
almost neutral in stability, the unstable growth of the mode with $q_c$
will generally be unable to lead to a Turing type periodic standing pattern.
Instead, from the outset, a group of modes with wavenumbers close to $q_c$
will couple nonlinearly with another group of modes of almost vanishing $q$ 
with almost neutral stability, leading to a peculiar spatio-temporal 
chaos\cite{Tribel,Dan}. 
The simple 
evolution equation
\begin{equation}
\partial_{t}u=-\partial_{x}^{2}[\epsilon-(1+\partial_{x}^{2})^{2}]u-(\partial
_{x}u)^{2},
\end{equation}
called the Nikolaevskii equation\cite{Nik}, gives qualitatively 
the same eigenvalue
spectrum as those in Fig.~6. Actually, it was argued previously that 
the Turing pattern in this system
existing for small positive $\epsilon$ is always unstable, and as a result
the system immediately becomes turbulent characterized by the coexistence of
turbulent fluctuations with vastly different length scales.
Possibility of this type of turbulence in reaction-diffusion systems
was discussed by Fujisaka and Yamada\cite{FY}.
In electro-convective systems, similar type of complex behavior was
discovered by Kai et al.\cite{Kai} which they called {\em soft-mode 
turbulence}.

\section{summary}
If spatially distributed limit cycle oscillators are diffusively coupled not 
directly but via
a certain diffusive component, the oscillators may be viewed as
coupled nonlocally after a mathematical elimination of this diffusive 
variable. We showed in the present paper that this actually 
occurs in reaction-diffusion
systems and that the nonlocality of this kind persists 
even close to
the Hopf bifurcation point provided the coupling between each  
local oscillator and the diffusive component is sufficiently weak.
Specifically, under this condition, the systems is reduced to a
comlex Ginzburg-Landau type equation with nonlocal rather than diffusive
coupling. 
Temporal nonlocality or memory effects which generally exist before
reduction does not appear explicitly in the reduced equation, still
they may generally affect the functional form of the coupling function.
Our results were generalized so as to include the effects of direct
but weak diffusive
coupling among the oscillators.

Linear stability analysis of the uniformly oscillating state of the
reduced equation was also carried out. Some new types of eigenvalues spectrum
were found to arise, and their physical relevance was suggested.

How the solution behaves in the nonlinear regime is known only partially.
For instance, type U$_{12}$ spectrum in Fig.~3 was found to lead to turbulence
with multi-affinity\cite{STC}. It was also found that even the normal 
spectrum (type S)
can give rise to peculiar spiral waves without phase singularity
when the nonlocal coupling becomes weak\cite{KS}. 
Different aspects of the nonlocal CGL including the afore-mentioned type
of turbulence associated with the dispersion curve in fig.~6 
will be developed in 
forthcoming papers\cite{Dan}.

\newpage
\begin{center}
\bf Figure Captions
\end{center}
\begin{itemize}
\item 
[Fig.1]  The sign of $u(q)$ changes with $q$ in three different ways. The
       figure shows how the corresponding domains A, B and C appear 
       in the $a_1$-$a_2$ plane.
       $u(q)$ is positive for all $q$ in A, negative above 
       some critical $q$ in B, and negative in a finite interval of $q$ in C. 
       The line separating domains A and C is given by a parabola 
       $a_2=a_1^2/4a_0$.
       
\item
[Fig.2]  The sign of $v(q)$ changes with $q$ in four different ways. The
       figure shows how the corresponding domains A$'$, B$'$, C$'$ and D$'$
       appear in the $b_0$-$b_1$ plane.
       $v(q)$ is positive for all $q$ in A$'$, negative above some critical 
       $q$ in B$'$, negative for all $q$ in C $'$, and 
       negative below some critical $q$ in D$'$.
       
\item
[Fig.3]  Four types of eigenvalue spectrum for the phase-like fluctuations
       about the uniform oscillation which is stable only for type S. 
       The figure shows how they appear 
       in the $K$-$Kc_1$ plane in the vicinity of the origin $K=0$.
       The other parameters are fixed as $c_2=1.0$ and $\theta=6.0$.
       In each of the four cases, the eigenvalues of the 
       amplitude-like fluctuations, which is not shown in the figure, form a
       branch completely separated from the phase branch and remain 
       negative for all $q$.
       
\item
[Fig.4]  Figure 3 is extended to the region of larger $K$. A new type of
       eigenvalue spectrum appears for large negative $K$. This type,
       labeled by U$_3$ is similar to U$_2$ in Fig.~3 except that 
       the 
       amplitude       
       branch merges with the phase branch above a certain wavenumber 
       which is still below the wavenumber at which $\rm{Re}\lambda$ 
       changes sign.
       Thus, the eigenvalues associated with 
       the unstable fluctuations are complex rather than real, i.e., the
       instability is oscillatory in nature.
       
\item
[Fig.5]  Similar to Fig.~4 but for different values of $c_2$ and $\theta$,
       i.e., $c_2=-2.5$ and $\theta=1.5$. New type of eigenvalue spectrum,
       denoted by U$_4$, appears which differs from U$_3$ in Fig.~4
       only in that the eigenvalues saturate to a negative value as $q$ goes
       to infinity.
\item       
[Fig.6]  Eigenvalue spectra close to a finite-wavenumber instability, which are
       obtained as a modification of U$_{2}$ in Fig.~3 by assuming 
       nonvanishing $\delta_1$ in the reduced equation (30).      
           
\end{itemize}
    
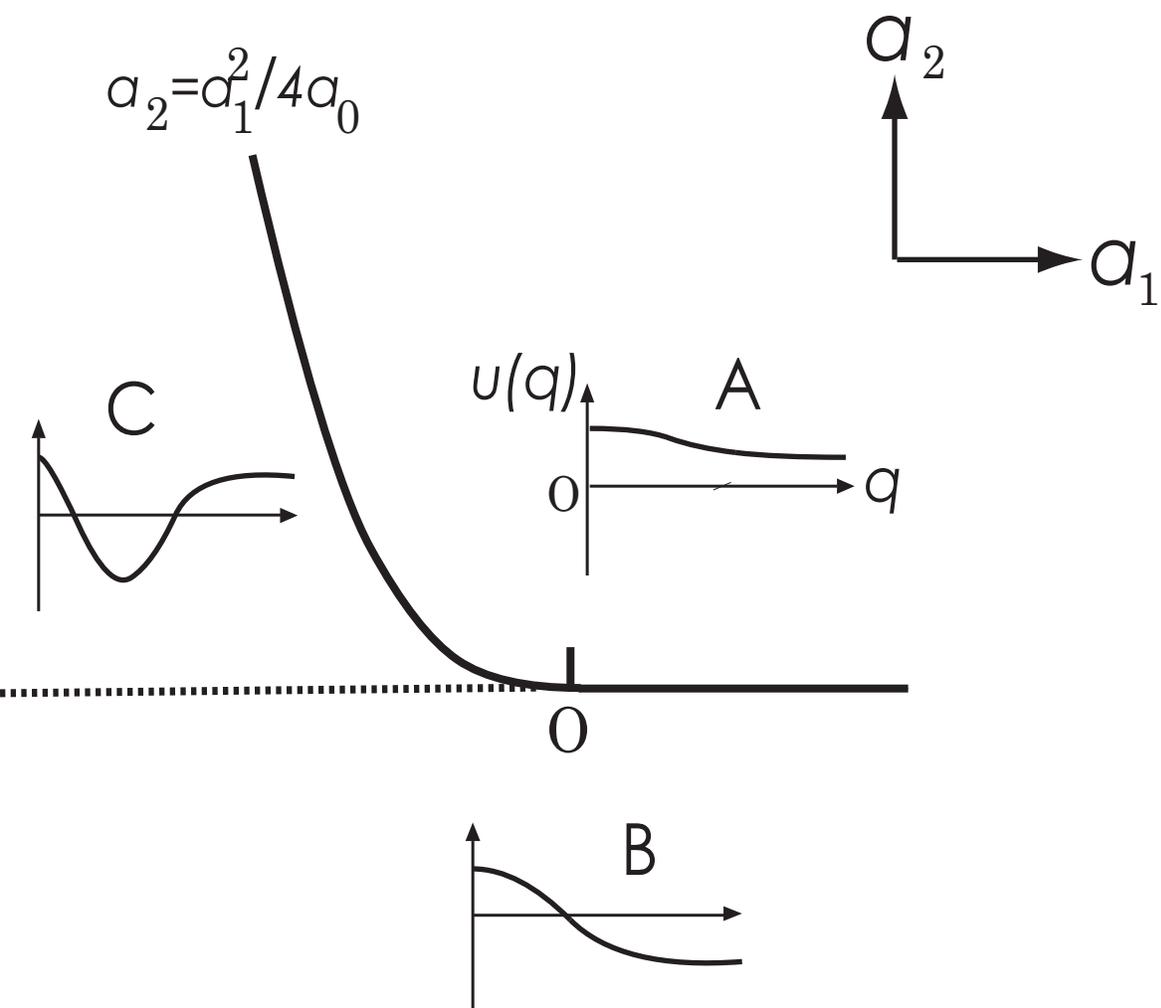
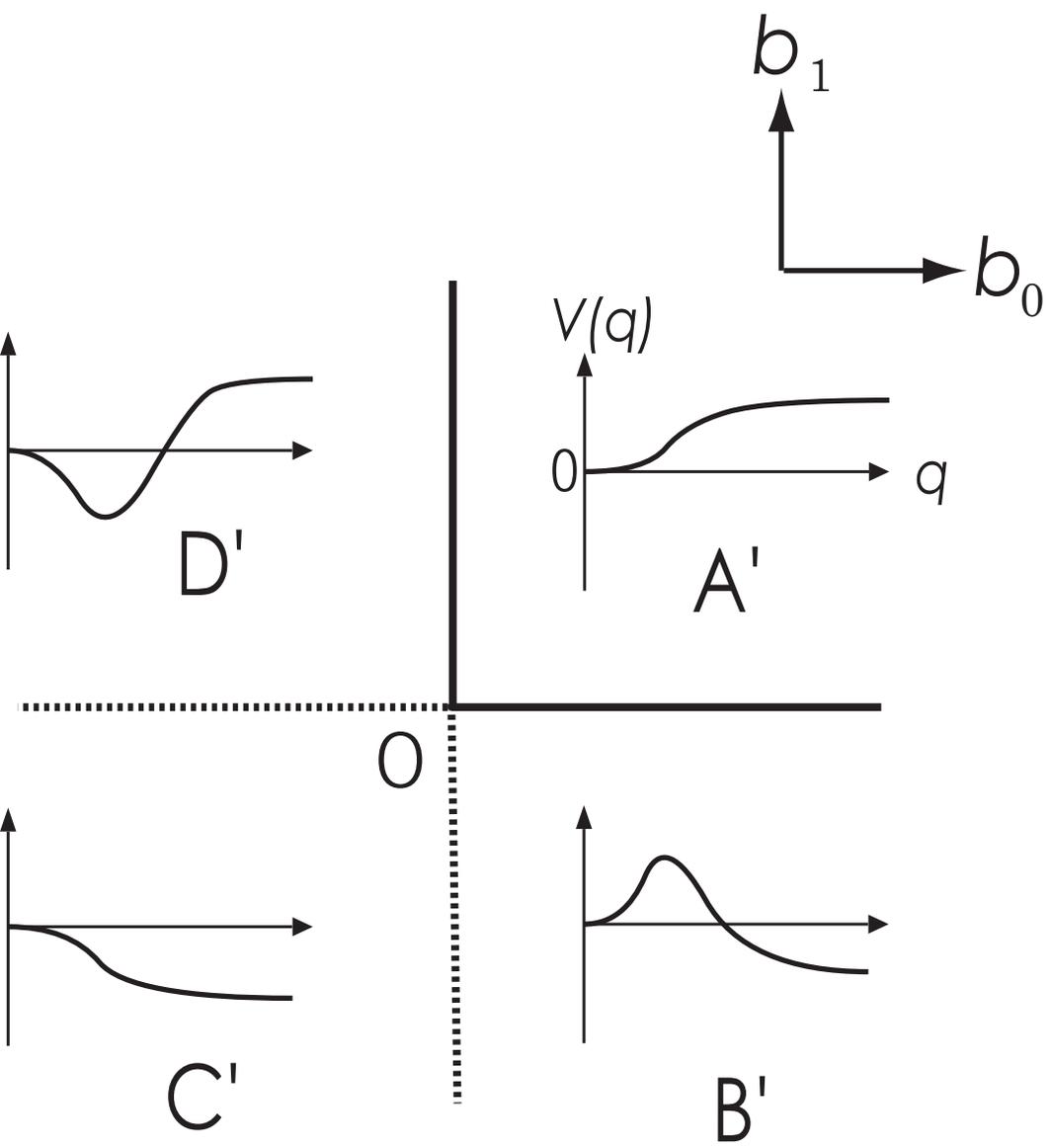
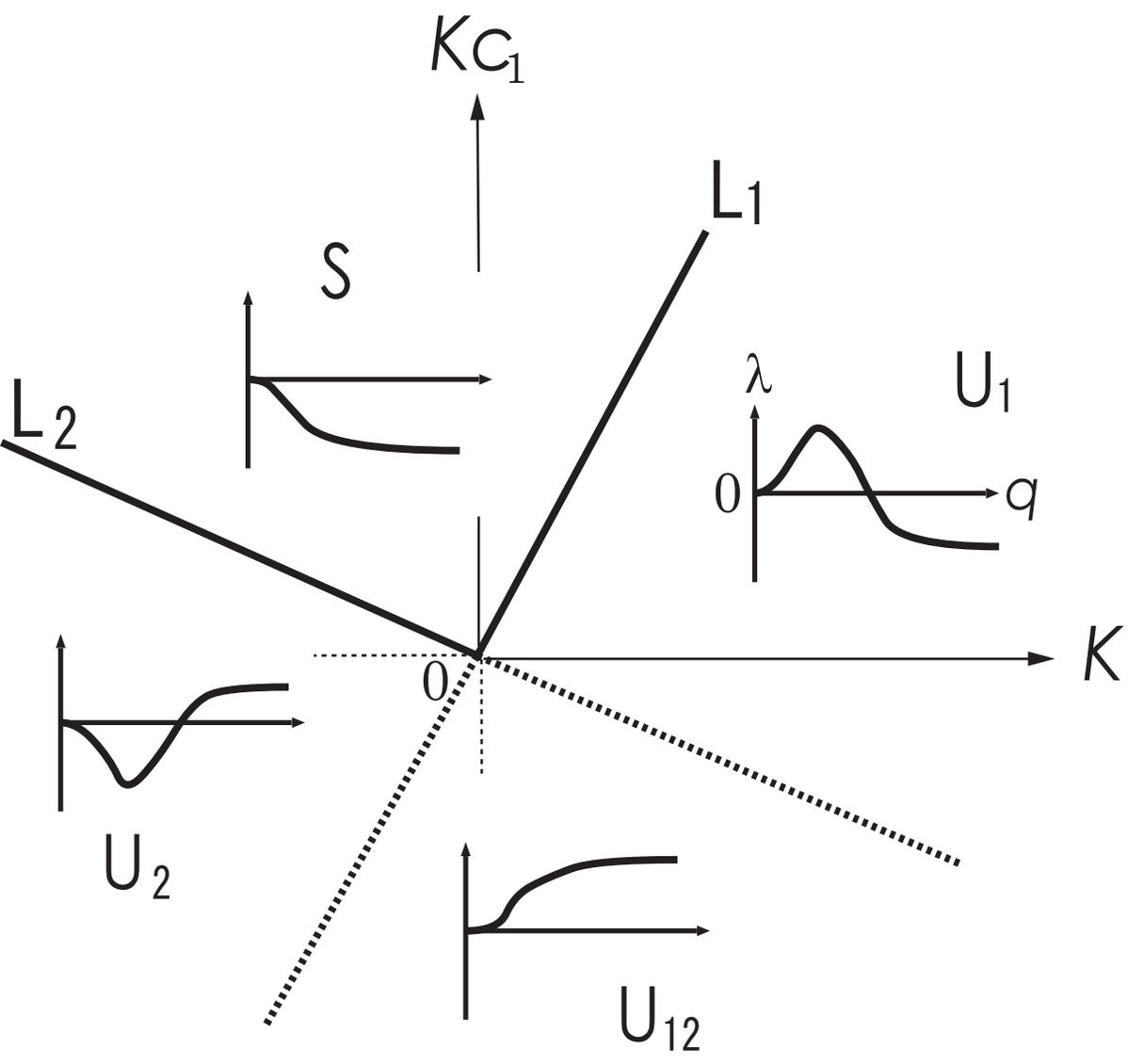
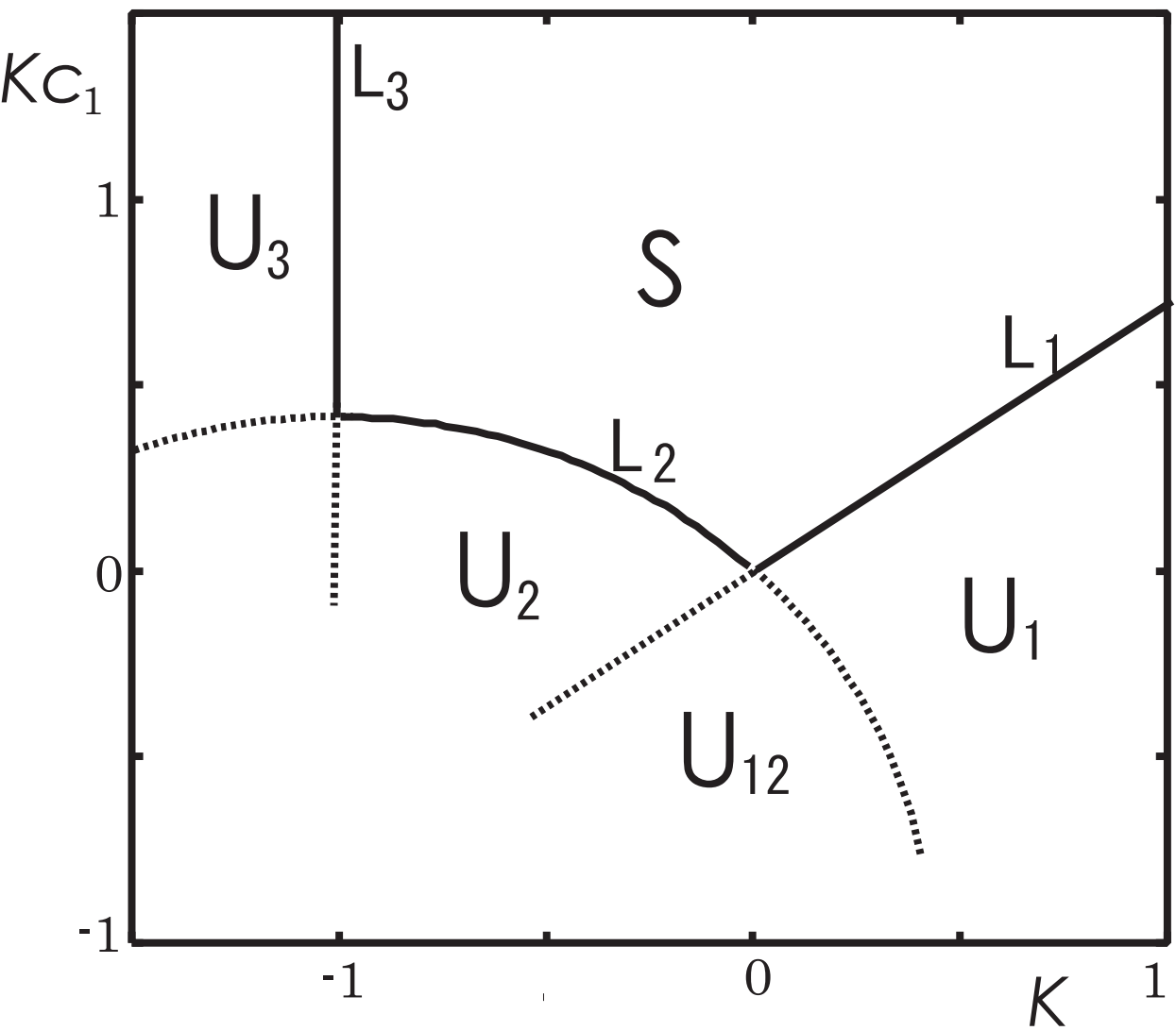
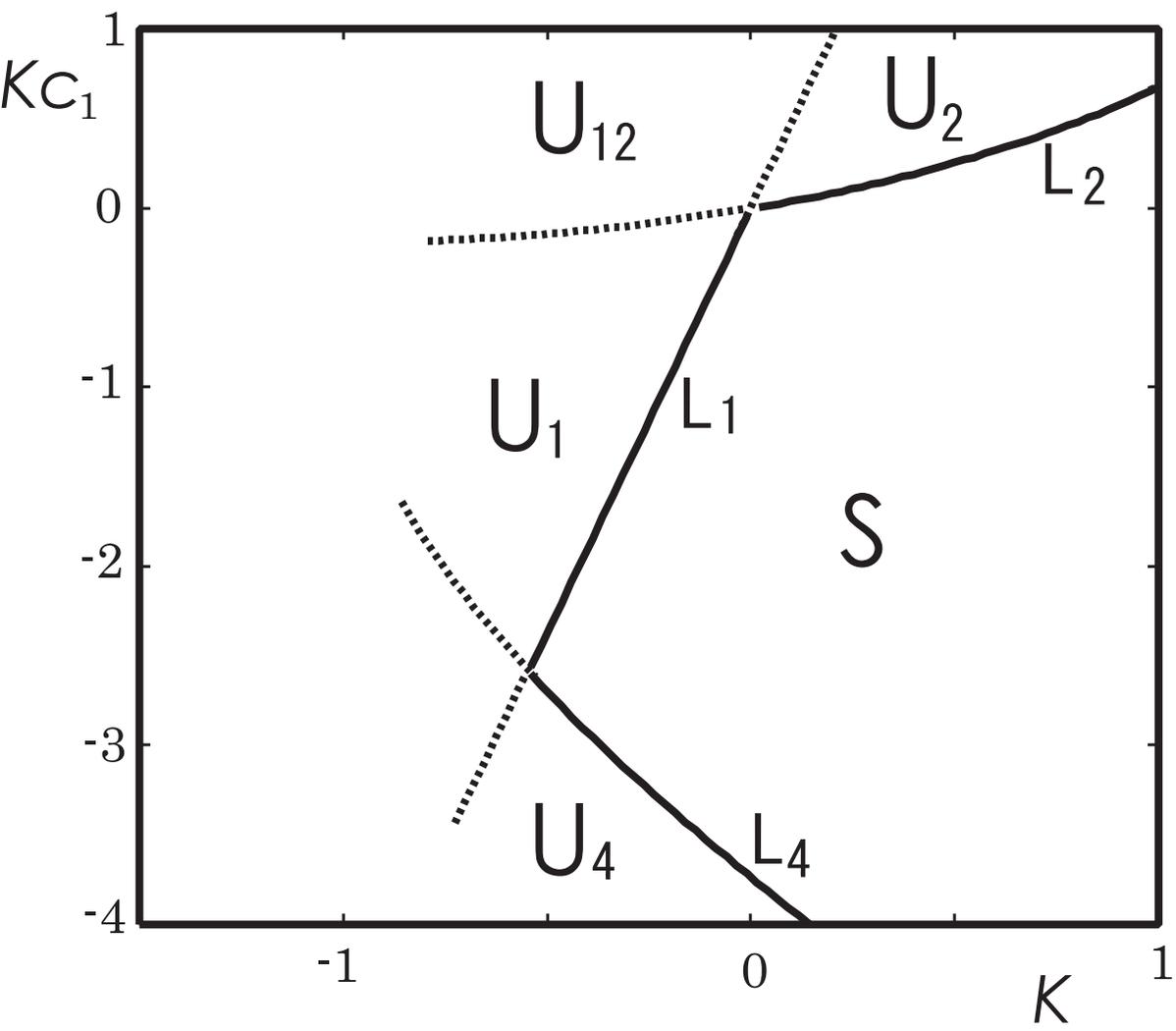
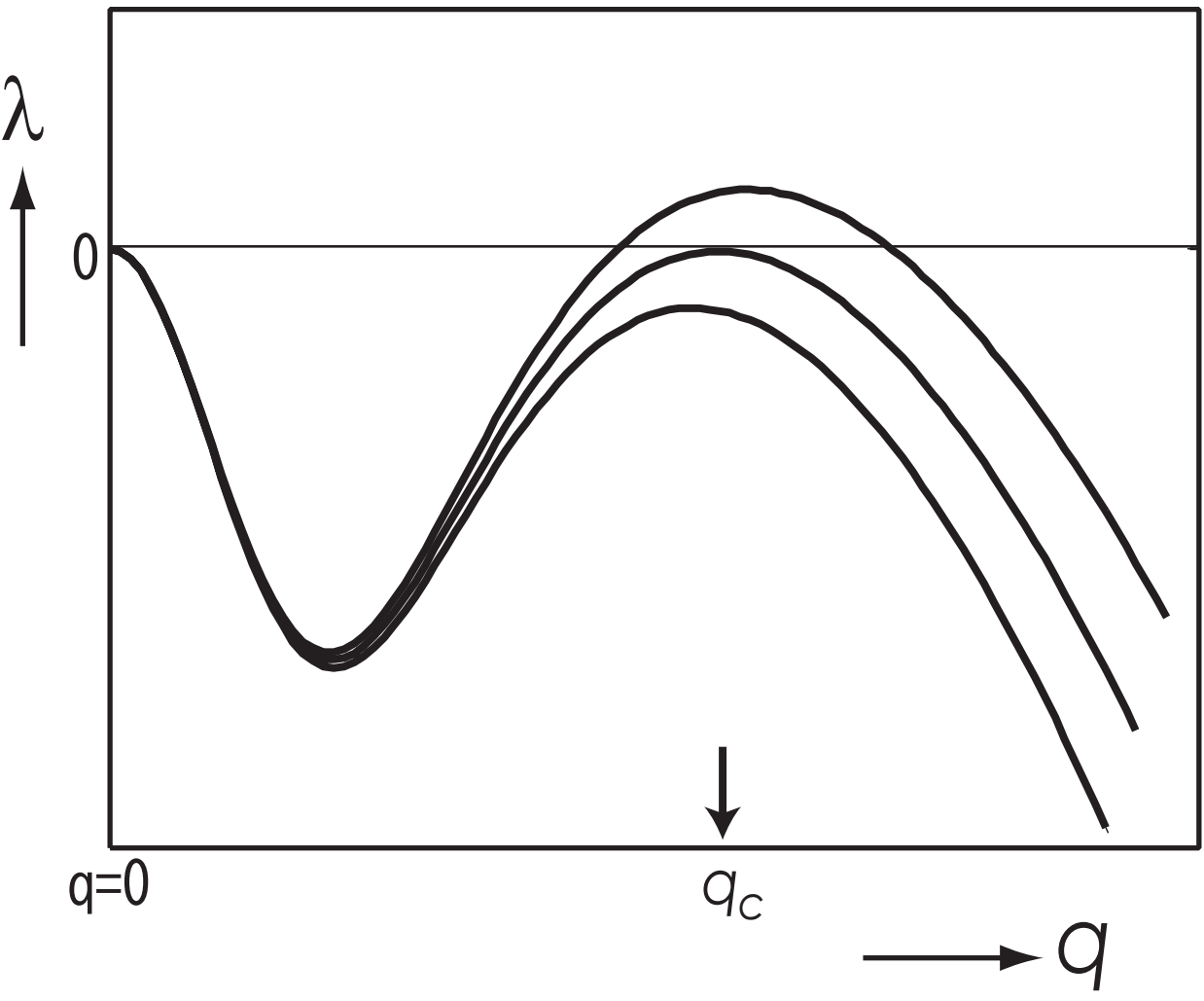
\end{document}